\documentclass{elsart}
\usepackage{graphicx}

\begin{document}
\begin{frontmatter}
\title{Pair dispersion in turbulence}

\author{G. Boffetta\thanksref{mail1}}
\address{Dipartimento di Fisica Generale
and INFM Unit\`a di Torino Universit\`a, \\
via Pietro Giuria 1, 10125 Torino, Italy}

and
\author{A. Celani}
\address{CNRS, Observatoire de la C\^ote d'Azur, B.P. 4229,
06304 Nice Cedex 4, France}

\thanks[mail1]{corresponding author, boffetta@to.infn.it}

\begin{abstract}
We study the statistics of pair dispersion in two-dimensional turbulence.
Direct numerical simulations show that
the pdf of pair separations is
in agreement with the Richardson prediction. 
The pdf of doubling times follows dimensional scaling
and shows a long tail which is the signature of close approaches
between particles initially seeded with a large separation.
This phenomenon is related to the formation of fronts in
passive scalar advection. 
\end{abstract}


\end{frontmatter}

\section{Introduction}
\label{sec:1}
The concentration of a dilute solution of a passive tracer in an
incompressible flow obeys the scalar equation  
\begin{equation}
\partial_t \theta + {\bf v} \cdot \nabla \theta = \kappa \Delta \theta + f
\label{1.1}
\end{equation}
where ${\bf v}$ is the divergenceless velocity field, $\kappa$ is the molecular
diffusivity, and $f$ is the external source of tracer fluctuations.
Equation (\ref{1.1}) can be solved by the method of characteristics to obtain
the solution
\begin{equation}
\begin{array}{lll}
\theta({\bf x},t)&=&\int_{-\infty}^{t} ds \, f({\bf \rho}(s),s) \\
\dot{{\bf \rho}}(s)&=&{\bf v}({\bf \rho}(s),s)+\sqrt{2\kappa}\,{\bf \eta}(s)\,, 
\qquad {\bf \rho}(t)={\bf x}  .
\end{array}
\label{1.2}
\end{equation}
The characteristics ${\bf \rho}(t)$ are the Lagrangian trajectories
of the fluid particles, and $\sqrt{2\kappa}\,{\bf \eta}$ is the white noise
contribution due to molecular diffusion.
There is an immense literature devoted to both Eulerian, (\ref{1.1}),
and Lagrangian, (\ref{1.2}), descriptions of passive transport in turbulence
\cite{MY75}. 
What is relevant to our purposes is to keep in mind the tight relationship
between these two complementary descriptions.
As an instance, simultaneous two-point correlations 
of the scalar field 
can be written in terms of two-particle Lagrangian statistics as
\begin{equation}
\langle \theta({\bf x}_1,t)\theta({\bf x}_2,t) \rangle=
\int_{-\infty}^{t} ds_1 \,
\int_{-\infty}^{t} ds_2 \,
\langle f({\bf \rho}(s_1),s_1)  f({\bf \rho}(s_2),s_2) \rangle \,.
\label{1.3}
\end{equation}
Properly choosing the form of the correlation function of the scalar forcing,
e.g. $\langle f({\bf x}_1,t_1)f({\bf x}_2,t_2) \rangle= \chi(|{\bf x}_1-{\bf x}_2|)\delta(t_1-t_2)$, 
and exploiting space homogeneity, 
expression (\ref{1.3}) can be further simplified to the form
\begin{equation}
\langle \theta({\bf x},t)\theta({\bf x}+{\bf R},t) \rangle=
\int_{-\infty}^{t} ds \, \int d{\bf r} \,
\chi({\bf r}) \,p({\bf r},s|{\bf R},t) \,.
\label{1.4}
\end{equation}
where $p({\bf r},s|{\bf R},t)$ is the probability density function for
a pair to be at a separation ${\bf r}$ at time $s$, under the condition that it has to have a separation ${\bf R}$ at time $t$. \\
Summarizing, the knowledge of the statistics of pair dispersion is sufficient to
determine the values of the correlations of passive scalar for a given forcing. 
On the contrary, to extract the Lagrangian statistics from the Eulerian one 
it is necessary to know the scalar correlation functions for different
forcings. In this sense, the Lagrangian information is more fundamental,
and hereafter we shall restrict to this one.

\section{Statistics of pair dispersion}
\label{sec:2}

The dispersion of a particles' pair in turbulence 
can be phenomenologically described in terms of a diffusion equation
for the probability distribution of pair separations
\begin{equation}
\frac{\partial p({\bf r},t)}{\partial t} =
\frac{\partial}{\partial r_j} \left(
K(r,t) \frac{\partial p({\bf r},t)}{\partial r_j} \right)
\label{eq:2.1}
\end{equation}
with a space and time dependent diffusion coefficient $K(r,t)$.
In general, the description by means of a diffusion equation is
a drastic simplification. The only case in which it can be proven 
that particle separations obey to (\ref{eq:2.1}) is when
the advecting velocity field is rapidly changing in time \cite{K68,BGK98}.

The original Richardson proposal, obtained from experimental data
in the atmosphere, is $K(r,t)=K(r) \sim r^{4/3}$ \cite{Richardson26}.
This leads to the well known non-Gaussian distribution
\begin{equation}
p({\bf r},t) \simeq t^{-9/2}
\exp\left(- C r^{2/3}/t \right)
\label{eq:2.2}
\end{equation}

From the Richardson distribution (\ref{eq:2.2}) one has immediately
that the mean square particle separation grows as
\begin{equation}
R^2(t) \equiv \langle r^2(t) \rangle \sim t^{3}
\label{eq:2.3}
\end{equation}
The ``$t^3$'' law, which is known as the Richardson law, has been 
observed, although with large uncertainty, in direct numerical simulations
\cite{ZB94} and, more recently, in laboratory experiments \cite{JPT99}.

The diffusion equation (\ref{eq:2.2}) is not 
the unique possibility which leads to the 
``$t^3$'' law. Batchelor \cite{Batchelor52} assuming that the diffusion
coefficient should depend on average quantities, proposed that 
$K(r,t)=K(t) \sim \langle r^2(t) \rangle^{2/3} \sim t^2$. 
In this case the distribution of pair distances is 
Gaussian with a superballistic growing variance
\begin{equation}
p({\bf r},t) \simeq t^{-9/2}
\exp\left(- C r^2/t^3 \right)
\label{eq:2.4}
\end{equation}

Of course, this is not the end of the story. Formally, any 
diffusion coefficient of the form $K(r,t) \sim r^{a} t^{b}$
with $3 a + 2 b = 4$ is compatible with the ``$t^3$'' law
but gives different distribution function \cite{MY75,KBS87}. 
Early experimental data were in favor of the Batchelor Gaussian
distribution (\ref{eq:2.4}) \cite{MY75}, but recent laboratory experiments
are more in the direction of the Richardson original proposal \cite{JPT99}.

The Richardson law can be derived by a simple dimensional argument
which makes use of the Kolmogorov similarity law for the Eulerian
velocity increments in fully developed turbulence. By definition
one has
\begin{equation}
{d \over dt} {1 \over 2} R^2(t) = 
\langle {\bf r} \cdot \delta{\bf v}^{(L)}({\bf r}) \rangle =
\langle r \delta v_{\parallel}^{(L)}(r) \rangle
\label{eq:2.5}
\end{equation}
where $\delta{\bf v}^{(L)}({\bf r})={\bf v}({\bf x}(t)+{\bf r})-
{\bf v}({\bf x}(t))$ is the Lagrangian velocity increment
and $\delta v_{\parallel}$ is its projection on the ${\bf r}$ 
direction.
Assuming Kolmogorov scaling for Lagrangian velocity differences,
$\delta v_{\parallel}(r) \sim r^{1/3}$, one obtains from
(\ref{eq:2.5}) the Richardson law (\ref{eq:2.3}).
The assumption that the Lagrangian velocity difference has the same
Kolmogorov scaling of the Eulerian one relies on the intuitive
idea that the main contribution to
the separation rate follows from eddies with size comparable
with the separation itself.

Let us conclude this section with a remark. The description of 
relative dispersion in terms of the diffusion equation 
(\ref{eq:2.1}) assumes a self-similarity in the process.
Of course this could not be the case, e.g. in intermittent
three dimensional fully developed turbulence.
As a matter of fact, in presence of intermittency of the velocity field,
one can expect a kind
of ``Lagrangian intermittency'', in the sense that different
moments $\langle r^p(t) \rangle$ have different scaling exponents:
\begin{equation}
\langle r^p(t) \rangle \sim t^{\alpha_p}
\label{eq:2.6}
\end{equation}
with $\alpha_p \neq 3/2 p$.
This problem has been discussed in several papers
\cite{GP84,CPV87,SWK87,Novikov89} with different conclusions.
Recent detailed investigations with a synthetic turbulent model
gave the evidence of Lagrangian intermittency with scaling
exponents $\alpha_p$ linked to the Eulerian intermittent
scaling exponents \cite{BCCV99}.

\section{Pair dispersion in two-dimensional turbulence}
\label{sec:3}

Pair dispersion statistics has been investigated by 
direct numerical simulation of the inverse energy cascade 
in two-dimensional turbulence. There are several reasons 
for considering 2D turbulence. From an applicative point
of view, two-dimensional Navier-Stokes equations are among
the simplest systems of geophysical interest. The 
observed absence of intermittency \cite{PT98,BCV99} makes 
the 2D inverse energy cascade an ideal framework for 
the study of Richardson scaling. Moreover, the dimensionality
of the problem makes feasible direct numerical simulations
at high Reynolds numbers.

The 2D Navier-Stokes equations for the vorticity
$\omega=\nabla \times {\bf v}= -\Delta \psi$ are:
\begin{equation}
\partial_t\omega+J\left(\omega,\psi\right)=\nu\Delta\omega
-\alpha\omega + \phi,
\label{eq:3.1}
\end{equation}
where $\psi$ is the stream function and $J$ denotes
the Jacobian. The friction linear term $-\alpha \omega$ extracts
energy from the system to avoid Bose-Einstein
condensation at the gravest modes \cite{SY93}. 
The forcing is active only on a typical scale $l_f$ and
is $\delta$-correlated in time to ensure the control of the
energy injection rate.
The viscous term has the role of removing enstrophy at scales smaller
than $l_f$ and, as customary, it is numerically more convenient to
substitute it by a hyperviscous term (of order eight in our
simulations).  
Numerical integration of (\ref{eq:3.1}) is performed by a standard
pseudospectral method on a doubly periodic square
domain of $N^2\!=\!2048^2$ grid points.
All the results presented 
are obtained in conditions of stationary turbulence.

\begin{figure}
\includegraphics[width=0.7\textwidth,height=0.3\textheight]{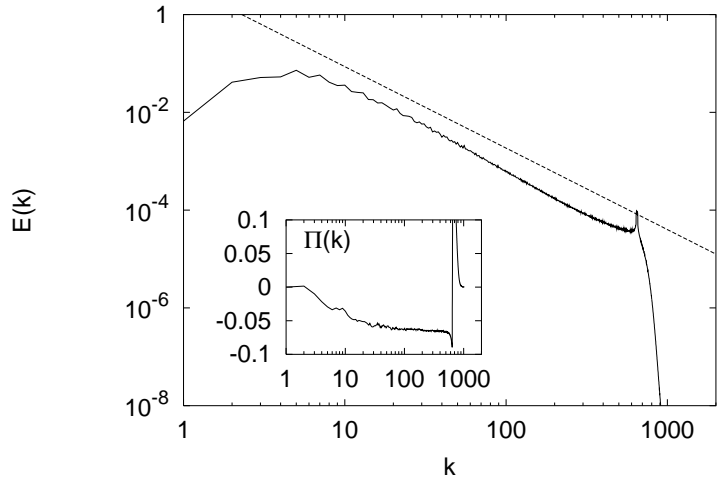}
\caption
{Energy spectrum $E(k)$. The dashed line is the Kolmogorov scaling 
$E(k) \simeq k^{-5/3}$. In the inset it is shown the energy flux $\Pi(k)$.}
\label{fig3.1}
\end{figure}

In Figure \ref{fig3.1} we present the energy spectrum, which displays
Kolmogorov scaling $E(k)=C \epsilon^{2/3} k^{-5/3}$ over about two decades
with Kolmogorov constant $C\simeq 6.0$. The
inertial range correspond to the region of constant flux,
also plotted in Figure \ref{fig3.1}.
Previous numerical investigation has shown that velocity differences
statistics in the inverse cascade is almost Gaussian with 
Kolmogorov scaling not affected by intermittency corrections \cite{BCV99}. 
In this case we expect the Lagrangian statistics to be self-similar
 with Richardson scaling \cite{BCCV99}.

\begin{figure}
\includegraphics[width=0.7\textwidth,height=0.3\textheight]{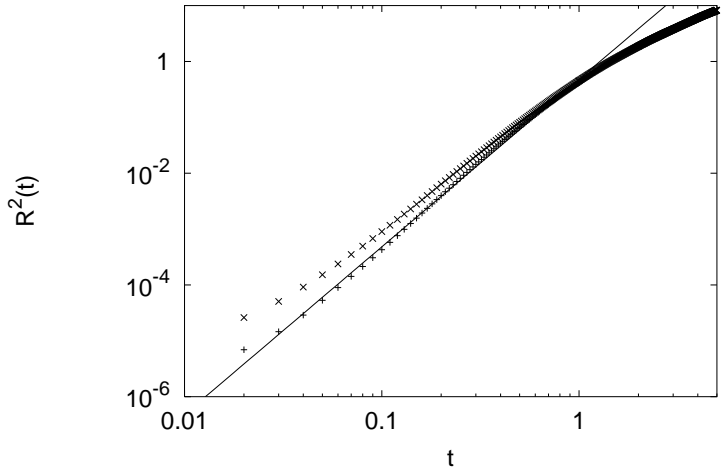}
\caption
{Relative dispersion $R^{2}(t)$ for two initial separation
$R(0)=1.5 \times 10^{-3}$ ($+$) and $R(0)=3 \times 10^{-3}$
($\times$). The continuous line is the Richardson law
$R^{2}(t) \simeq t^3$.}
\label{fig3.2}
\end{figure}

In Figure \ref{fig3.2} we plot relative dispersion $R^2(t)$ obtained 
after averaging over $64000$ particle pairs for two different 
initial conditions $R^2(0)$. The Richardson $t^3$ law is observed in
a limited time interval,
especially for the largest $R^2(0)$ run. It is remarkable that
the relative separation law displays such a strong dependence
on the initial conditions even in this high resolution runs.

\begin{figure}
\includegraphics[width=0.7\textwidth,height=0.3\textheight]{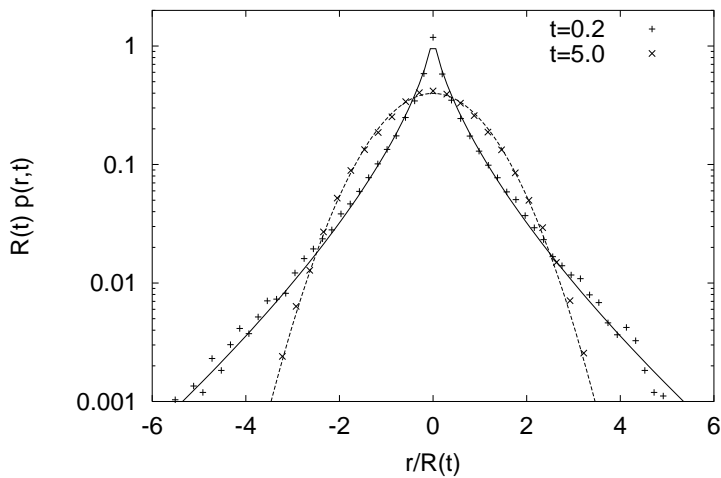}
\caption{
Probability density function of relative separation at times 
$t=0.2$ and $t=5.0$ rescaled with $R(t)=\langle r^{2}(t) \rangle^{1/2}$.
The continuous line is the Richardson prediction (\ref{eq:2.2}),
the dashed line is a Gaussian distribution.
}
\label{fig3.3}
\end{figure}

The probability density functions of pair separations is plotted
in Figure \ref{fig3.3} at two different times. For short times
($t=0.2$) in which the relative dispersion is in the Richardson 
regime (see Figure \ref{fig3.2}) we see that the Richardson 
distribution (\ref{eq:2.2}) fits well the numerical data. 
This is, we think, a clear evidence of the substantial validity
of the original Richardson description. Let us observe that
until now this point was not clear: recent laboratory data \cite{JPT99}
pointed to the Richardson distribution 
but there were strong deviations from (\ref{eq:2.2}).
At long time $t=5.0$ relative separation distribution is described
by a Gaussian distribution, but this has no relation with the
Batchelor proposal (\ref{eq:2.4}) because it is not in the scaling
range. At time $t=5$ the average separation is of the size of
the computational box and we have normal dispersion {\it \`a la} Taylor.

The Richardson pdf has a strong cusp at $r=0$ which signals the high
probability for a pair to reach a very small separation compared to the
typical value $R(t)$. As we shall see later on, this effect is highlighted
by considering the statistics of first exit-times.

\section{Doubling time statistics}
\label{sec:4}

We have seen in the previous section that the Richardson 
picture seems to be confirmed in the inverse energy cascade 
in two dimensions.
Nevertheless we have seen that it is difficult to observe
the Richardson scaling law even in our high resolution
direct numerical simulations. 
To understand this effect consider a series of particle pair dispersion
experiments, in which a couple of particles is released
at time $t=0$ with initial separation $R(0)$.
At a {\it fixed time} $t$ one performs an average over all different
experiments and computes $R^2(t)$.
It is clear that, unless $t$ is large enough that all particle pairs have
``forgotten'' their initial conditions, the average will be biased.
This is at the origin of the flattening of $R^2(t)$
for small times, which we can call a crossover from initial
condition to self similar regime. Of course, at larger $R(0)$
correspond longer crossover regimes (see Figure \ref{fig3.2}).
A similar effect is observed for times
of the order of the integral time-scale since some particle
pairs might have
reached a separation larger than the integral scale and thus
diffuse normally, biasing the average, so that the curve
$R^2(t)$ flattens again.

To overcome this difficulty we use an alternative approach
based on statistics at {\it fixed scale} \cite{ABCCV97}.
The method has been successfully applied to 
the analysis of relative dispersion in synthetic turbulent
flow \cite{BCCV99} and in experimental convective laminar
flow \cite{BCEQ99}. The method works as follows.
Given a set of thresholds $R_n=r^{n} R(0)$ within the inertial range,
one computes the ``doubling time'' $T_{r}(R_n)$ defined as the time it takes
for the particle pair separation to grow from threshold $R_n$ to
the next one $R_{n+1}$.
Averages are then performed over many dispersion experiments, i.e.,
particle pairs.
The outstanding advantage of this kind of averaging at fixed scale separation,
as opposite to a fixed time,
is that it removes crossover effects since all sampled particle pairs
belong to the inertial range.

The scaling property of the doubling time statistics in fully developed
turbulence is obtained by a simple dimensional argument. 
The time it takes for the particle pair separation
to grow from $R$ to $r R$ can be dimensionally estimated as
$T_{r}(R) \sim R/\delta v_{\parallel}^{(L)}(R)$;
we thus expect for the inverse doubling times the scaling
$\langle T_{r}(R) \rangle \simeq R^{2/3}$.
From the definition, doubling times depend on the threshold
ratio $r$. It is then useful to consider the normalized quantity
\begin{equation}
\lambda(R) = {1 \over \langle T(R) \rangle} \log r
\label{eq:4.1}
\end{equation}
which is called the Finite Size Lyapunov Exponent (FSLE) 
because is reduces to the (Lagrangian) Lyapunov exponent 
in the limit of small separation $R \rightarrow 0$ \cite{ABCPV97}.

\begin{figure}
\includegraphics[width=0.7\textwidth,height=0.3\textheight]{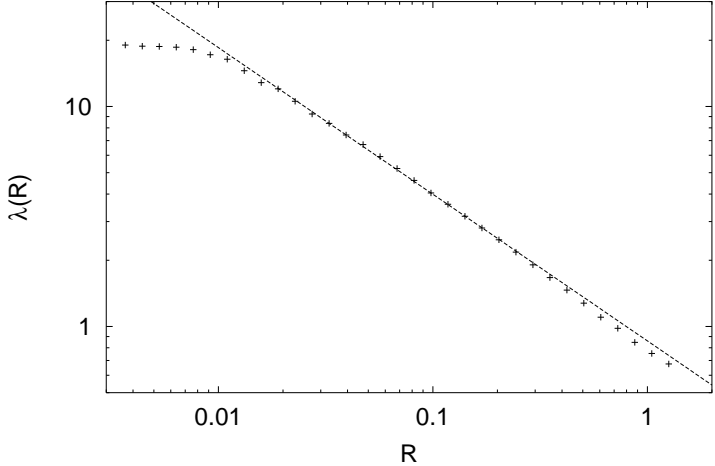}
\caption{
Finite Size Lyapunov Exponent (\ref{eq:4.1}) for 
the same trajectories of Figure \ref{fig3.2}. The initial 
threshold is $R(0)=0.0031$ and the ratio is $r=1.2$.
The line is the theoretical Richardson scaling $R^{-2/3}$.
}
\label{fig4.1}
\end{figure}

In Figure \ref{fig4.1} it is shown the FSLE for the same simulation
of Figure \ref{fig3.2}. At small scales $R<0.01$ there appears a
constant plateau corresponding to the Lagrangian Lyapunov exponent
$\lambda(R) \simeq 19$. At larger $R$, we observe the power law 
$\lambda(R) \simeq R^{-2/3}$ on a scaling range which is well enhanced 
with respect to the relative dispersion of Figure \ref{fig3.2}.

The scaling of doubling times gives also information about
the two-point correlations of passive scalar.
Indeed, assuming that the correlation of the scalar forcing  
decays rapidly to zero with a typical scale $L$,
by means of equation (\ref{1.4}) we obtain 
\begin{equation}
\langle \theta({\bf x},t)\theta({\bf x}+{\bf R},t) \rangle \simeq
\int_{-\infty}^{t} ds \, \int_{|{\bf r}|<L} d{\bf r} \,
p({\bf r},s|{\bf R},t) 
\end{equation}
that is the average time that a particle pair released 
at a separation $R$ spends below the scale $L$.
Thus, our results for doubling times enable us to express
the scalar correlation as 
\begin{equation}
\langle \theta({\bf x},t)\theta({\bf x}+{\bf R},t) \rangle\sim
L^{2/3}-R^{2/3}
\end{equation}
for any $L$ and $R$ in the scaling range of the velocity field.
Direct numerical simulations of passive scalar advection, eq. (\ref{1.1}),
in the Navier-Stokes flow generated by eq. (\ref{eq:3.1}) confirm
that the exponent of scalar correlations is indeed indistinguishable
from $2/3$ \cite{CLMV99}.

Beyond scaling properties of averaged quantities, the inspection
of the pdf of doubling times is very insightful to capture the
main features of pair dispersion.

\begin{figure}
\includegraphics[width=0.7\textwidth,height=0.3\textheight]{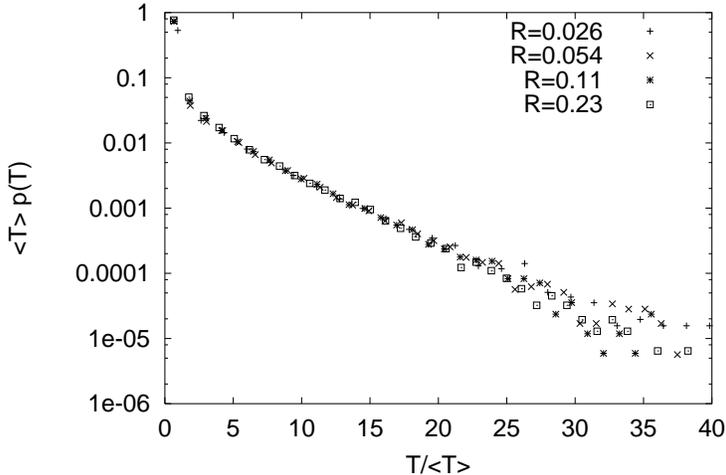}
\caption{
Probability distribution functions of doubling times rescaled with the
average doubling time $\langle T(R) \rangle$ for different $R$ in the
inertial range. 
}
\label{fig4.2}
\end{figure}

In figure \ref{fig4.2} it is shown the pdf of doubling times rescaled
with their average values. 
The normalized pdf's at different separations in the inertial range
collapse, indicating the self-similarity of the Lagrangian 
dispersion. Most important, there is a large number of events for which
the pair wanders for $20-30$  times the average value before
exiting. This effect is a reflection of the strong cusp observed in the
pdf of pair separations. In these events the particles which are initially
at a separation lying well inside the inertial range can approach each other
as close as the diffusive scale.
In the language of the passive scalar field, 
since trajectories originating from widely separated regions of space
can carry very different values of the concentration field $\theta$,
these approaches generate steep gradients of scalar across small scales.
These structures, known as ``cliffs'', 
have been actually observed both experimentally \cite{CG77,PGM82,KRS91}
and numerically \cite{HS94,AP94}
for the temperature field in a turbulent flow.

\section{Conclusion}
Passive scalar transport in turbulence can be described in
two complementary ways. It is possible to adopt the field description,
and think in terms of correlation functions of the scalar field,
either to prefer the particle description, and thus ask questions about 
the statistics of relative separations. These two aspects
complete each other. We have investigated the Lagrangian properties
of transport in two-dimensional turbulence.
our results show that pair dispersion statistics is not intermittent,
since the velocity field is self-similar and the geometric content
of two-particle configurations is trivial. This result makes contact
with the analytical result that can be derived in the rapid-change model,
where two-point scalar correlations show no anomaly.
This will not be the case in three-dimensional turbulence, a case which is
under current analysis.\\
Although on average particle pairs separate, 
probability density functions
of pair separations  and of doubling times clearly display the fingerprint
of frequent close approaches between particles. 
These events occur with a relatively 
high probability, and are for the formation of
quasi-discontinuities in the scalar field (the cliffs). 

\begin{ack}
We thank M.~Vergassola for useful discussions. 
The support by INFM (Progetto di Ricerca Avanzata TURBO)
is acknowledged. The numerical simulation were perfomed at
CINECA under the INFM contract ``Lagrangian and Eulerian
statistics in fully developed turbulence''.
\end{ack}



\begin{thebibliography}{99}

\bibitem{MY75}
A.~Monin and A.~Yaglom,
Statistical Fluid Mechanics,
(MIT Press, Cambridge, Mass., 1975).

\bibitem{K68}
R.~H.~Kraichnan, 
Phys. Fluids. 11 (1968) 945.

\bibitem{BGK98}
D.~Bernard, K.~Gawedzki and A.~Kupiainen,
J. Stat. Phys. 90 (1998) 519.

\bibitem{Richardson26}
L.~F.~Richardson, Proc. Roy. Soc. A110 (1926) 709.

\bibitem{ZB94}
N.~Zovari and A.~Babiano,
Physica D 76 (1994) 318.

\bibitem{JPT99}
M.~C.~Jullien, J.~Paret and P.~Tabeling,
Phys. Rev. Lett. 82 (1999) 2872.

\bibitem{Batchelor52}
G.~K.~Batchelor, Proc. Camb. Phil. Soc. 48 (1952) 345.

\bibitem{KBS87}
J.~Klafter, A.~Blumen and M.~F.~Shlesinger,
Phys. Rev. A 35 (1987) 3081.

\bibitem{GP84}
S.~Grossmann and I.~Procaccia,
Phys. Rev. A 29 (1984) 1358.

\bibitem{CPV87}
A.~Crisanti, G.~Paladin and A.~Vulpiani,
Phys. Lett. A 126 (1987) 120.

\bibitem{SWK87}
M.~F.~Shlesinger, B.~J.~West and J.~Klafter,
Phys. Rev. Lett. 58 (1987) 1100.

\bibitem{Novikov89} 
E.~A.~Novikov,
Phys. Fluids A 1 (1989) 326.

\bibitem{BCCV99}
G.~Boffetta, A.~Celani, A,~Crisanti and A.~Vulpiani,
Europhys. Lett. 46 (1999) 177.

\bibitem{PT98}
J.~Paret and P.~Tabeling, 
Phys. Fluids 10 (1998) 3126.

\bibitem{BCV99}
G.~Boffetta, A.~Celani and M.~Vergassola,
Phys. Rev. E (in press, 1999).

\bibitem{SY93}
L.~Smith and V.~Yakhot, 
Phys. Rev. Lett. 71 (1993) 352.

\bibitem{ABCCV97}
V.~Artale, G.~Boffetta, A.~Celani, M.~Cencini and A.~Vulpiani,
Phys. Fluids A 9 (1997) 3162.

\bibitem{BCEQ99}
G.~Boffetta, M.~Cencini, S.~Espa and G.~Querzoli,
Europhys. Lett. 48 (1999) 629.

\bibitem{ABCPV97}
E.~Aurell, G.~Boffetta, A.~Crisanti, G.~Paladin and A.~Vulpiani
J. Phys. A 30 (1997) 1.

\bibitem{CLMV99}
A.~Celani, A.~Lanotte, A.~Mazzino and M.~Vergassola,
chao-dyn/9909038

\bibitem{CG77} 
C.~Gibson, C.~Freihe and S. McConnell,  Phys. Fluids 20 (1977) S156.

\bibitem{PGM82}
P.G.~Mestayer, J. Fluid Mech. 125 (1982) 475.

\bibitem{KRS91}
K.R.~Sreenivasan,  Proc. Roy. Soc. London  A434 (1991) 165.

\bibitem{HS94}
M.~Holzer and E.~Siggia, Phys. Fluids  6 (1994) 1820.

\bibitem{AP94}
A.~Pumir, Phys. Fluids 6 (1994) 2118.

\end{thebibliography}
\end{document}